\documentclass[amsmath,amssymb,showpacs,twocolumn]{revtex4}
\usepackage{graphicx}
\usepackage{amscd,amsmath,amsthm,amsfonts,amssymb}

\def\tanh{{\rm tanh}}
\def\Imag{{\rm Im}}
\def\Real{{\rm Re}}

\def\[{\begin{equation}}
\def\]{\end{equation}}

\begin{document}
\title{Universal patterns of rogue waves}
\author{Bo Yang and Jianke Yang}

\address{Department of Mathematics and Statistics, University of Vermont, Burlington, VT 05405, USA}

\begin{abstract}
Rogue wave patterns in the nonlinear Schr\"{o}dinger (NLS) equation and the derivative NLS equation are analytically studied. It is shown that when the free parameters in the analytical expressions of these rogue waves are large, these waves would exhibit the same patterns, comprising fundamental rogue waves forming clear geometric structures such as triangle, pentagon, heptagon and nonagon, with a possible lower-order rogue wave at its center. These rogue patterns are analytically determined by the root structures of the Yablonskii-Vorob'ev polynomial hierarchy, and their orientations are controlled by the phase of the large free parameter. This connection of rogue wave patterns to the root structures of the Yablonskii-Vorob'ev polynomial hierarchy goes beyond the NLS and derivative NLS equations, and it gives rise to universal rogue wave patterns in integrable systems.
\end{abstract}


\maketitle

Rogue waves are large and spontaneous nonlinear wave excitations that ``appear from nowhere and disappear with no trace" \cite{Akhmediev1}. They are a threat to ships in the ocean and cause various extreme events in optical systems, and have thus received intensive studies in recent years (see \cite{Ocean_rogue_review,Pelinovsky_book,Solli_Nature,Wabnitz_book} for reviews). So far, analytical expressions of rogue waves have been derived in a wide array of integrable physical models, such as the nonlinear Schr\"{o}dinger (NLS) equation for wave-packet propagation in the ocean and optical systems \cite{Benney,Peregrine,AAS2009,DGKM2010,KAAN2011,GLML2012,OhtaJY2012}, the derivative NLS equation for circularly polarized nonlinear Alfv\'en waves in plasmas \cite{Kaup_Newell,KN_Alfven1,KN_rogue_2011,KN_rogue_2013,YangDNLS2019}, and the Manakov equations for light transmission in randomly birefringent fibers \cite{Menyuk,ManakovDark}. Some of the predicted rogue wave solutions have also been observed in both water-wave and optics experiments \cite{Tank1,Tank2,Fiber1,Fiber2}.

The study of rogue wave patterns is important as it allows for the prediction of later rogue wave events from earlier wave forms. Although rogue wave solutions have been obtained in many physical integrable equations, and low-order rogue wave graphs in those systems have been plotted,  systematic studies of rogue wave patterns, especially the richer patterns arising from high-order rogue wave solutions, is still very limited.
For the NLS equation, preliminary investigations on rogue patterns were reported in \cite{KAAN2011,HeFokas,KAAN2013} through Darboux transformation and numerical simulations. It was observed in \cite{KAAN2011} that if a $N$-th order rogue wave exhibits a ring structure, then the center of the ring is a $(N-2)$-th order rogue wave. This observation was explained analytically in \cite{HeFokas}. In \cite{KAAN2013}, it was observed that NLS rogue patterns could be classified according to the order of the rogue waves and the parameter shifts applied to the Akhmediev breathers in the rogue-wave limit. This latter observation allowed the authors to extrapolate the shapes of rogue waves beyond order six. Despite these investigations, analytical and quantitative predictions of NLS rogue patterns at arbitrary orders are still nonexistent. It is also unclear if there is any connection between rogue wave patterns in the NLS equation and those in other integrable systems.

In this article, we reveal the deep connection between rogue wave patterns and root structures of the Yablonskii-Vorob'ev polynomial hierarchy. We show that when the free parameters in the analytical expressions of these rogue waves are large, these waves would exhibit clear geometric patterns, which are formed by fundamental rogue waves arranged in shapes such as triangle, pentagon, heptagon and nonagon, with a possible lower-order rogue wave at its center. The geometric shapes of these rogue patterns are analytically predicted by the root structures of the Yablonskii-Vorob'ev polynomial hierarchy, and their orientations are controlled by the phase of the large free parameter.
Although these results are explicitly derived only for the NLS and derivative NLS equations, they are valid for other integrable equations as well. Thus, universality of rogue wave patterns through root structures of the Yablonskii-Vorob'ev polynomial hierarchy is established. These universal rogue patterns deepen our understanding of rogue wave phenomenon, and make prediction of sophisticated rogue events possible in relevant physical systems.

First, we consider rogue patterns in the NLS equation
\[ \label{NLS-2020}
\textrm{i} u_{t} + \frac{1}{2}u_{xx}+ |u|^2 u=0.
\]
Analytical expressions for general rogue waves in this equation have been derived in \cite{DGKM2010,GLML2012,OhtaJY2012} by various methods. However, those expressions are not the best for solution analysis. Here, we present a simpler expression for these solutions, which can be readily derived by incorporating a new parameterization \cite{YangDNLS2019} into bilinear rogue waves in \cite{OhtaJY2012}. These simpler expressions, for rogue waves with unit-amplitude boundary conditions of $u(x,t)\to e^{\textrm{i}t}$ as $x, t\to \pm \infty$, are
\begin{eqnarray}
&& u_N(x,t)=\frac{\sigma_{1}}{\sigma_{0}}e^{\textrm{i}t}, \label{BilinearTrans2}
\end{eqnarray}
where the positive integer $N$ represents the order of the rogue wave,
$\sigma_{n}$ is a $N \times N$ Gram determinant
\begin{equation} \label{sigma_n}
\sigma_{n}=
\det_{
\begin{subarray}{l}
1\leq i, j \leq N
\end{subarray}
}
\left(
\begin{array}{c}
 m_{2i-1,2j-1}^{(n)}
\end{array}
\right),
\end{equation}
the matrix elements in $\sigma_{n}$ are defined by
\begin{equation*}
m_{i,j}^{(n)}=\sum_{\nu=0}^{\min(i,j)} \frac{1}{4^{\nu}} \hspace{0.06cm} S_{i-\nu}(\textbf{\emph{x}}^{+}(n) +\nu \textbf{\emph{s}})  \hspace{0.06cm} S_{j-\nu}(\textbf{\emph{x}}^{-}(n) + \nu \textbf{\emph{s}}),
\end{equation*}
vectors $\textbf{\emph{x}}^{\pm}(n)=\left( x_{1}^{\pm}, x_{2}^{\pm},\cdots \right)$ are defined by
\begin{eqnarray*}
&&x_{1}^{\pm}=x \pm \textrm{i} t \pm n, \ \ \ x_{2k}^{\pm} = 0, \\
&&x_{2k+1}^{+}= \frac{x+2^{2k} (\textrm{i} t)}{(2k+1)!} +a_{2k+1},    \\
&&x_{2k+1}^{-}=  \frac{x-2^{2k} (\textrm{i} t)}{(2k+1)!}+ a_{2k+1}^*,
\end{eqnarray*}
with the asterisk * representing complex conjugation,
$\textbf{\emph{s}}=(s_1, s_2, \cdots)$ are coefficients from the expansion
\begin{eqnarray*}
\sum_{r=1}^{\infty} s_{r}\lambda^{r}=\ln \left[\frac{2}{\lambda}  \tanh \left(\frac{\lambda}{2}\right)\right],
\end{eqnarray*}
the Schur polynomials $S_k(\mbox{\boldmath $x$})$, with $\emph{\textbf{x}}=\left( x_{1}, x_{2}, \ldots \right)$, are defined by
\begin{equation} \label{Schurdef}
\sum_{k=0}^{\infty}S_k(\mbox{\boldmath $x$}) \epsilon^k
=\exp\left(\sum_{k=1}^{\infty}x_k \epsilon^k\right),
\end{equation}
and $a_{2k+1} \hspace{0.05cm} (k=0, 1,\cdots, N-1)$ are free complex constants. Of these $N$ free complex constants, we will normalize $a_1=0$ by a shift of the $x$ and $t$ axis. Thus, the above general rogue waves have $N-1$ free irreducible complex parameters $a_3, a_5, \cdots, a_{2N-1}$.

Clear and recognizable rogue wave patterns will emerge when some of these $N-1$ free parameters get large. In this article, we will analytically determine these rogue patterns when only one of these free parameters is large, while the other parameters remain $O(1)$. It turns out that the resulting rogue patterns are completely determined by the Yablonskii-Vorob'ev polynomial hierarchy, and this polynomial hierarchy will be introduced first.

Yablonskii-Vorob'ev polynomials arose in rational solutions of the second Painlev\'{e} equation \cite{Yablonskii1959,Vorobev1965}, and
their determinant expressions were derived in \cite{Kajiwara-Ohta1996}. Let $p_{k}(z)$ be the special Schur polynomial defined by
\begin{equation*}
\sum_{k=0}^{\infty}p_k(z) \epsilon^k =\exp\left( z \epsilon - \frac{4}{3}\epsilon^3 \right).
\end{equation*}
Then, Yablonskii-Vorob'ev polynomials $Q_{N}(z)$ are given by a $N \times N$ determinant \cite{Kajiwara-Ohta1996,Clarkson2003-II}
\begin{eqnarray*}
&& Q_{N}(z) = \gamma_{N} \left| \begin{array}{cccc}
         p_{N}(z) & p_{N+1}(z) & \cdots &  p_{2N-1}(z) \\
         p_{N-2}(z) & p_{N-1}(z) & \cdots &  p_{2N-3}(z) \\
        \vdots& \vdots & \vdots & \vdots \\
         p_{-N+2}(z) & p_{-N+3}(z) & \cdots &  p_{1}(z)
       \end{array}
 \right|,
\end{eqnarray*}
where $\gamma_{N}= \prod_{j=1}^{N}(2j-1)!!$, and $p_{k}(z)= 0$ if $k<0$. To define the Yablonskii-Vorob'ev polynomial hierarchy, we let $p_{k}^{[m]}(z)$ be the generalized Schur polynomial defined by
\begin{equation} \label{pkmz}
\sum_{k=0}^{\infty}p_k^{[m]}(z) \epsilon^k =\exp\left( z \epsilon - \frac{2^{2m}}{2m+1}\epsilon^{2m+1} \right),
\end{equation}
where $m$ is a positive integer. Then, the Yablonskii-Vorob'ev hierarchy $Q_{N}^{[m]}(z)$ are given by the $N \times N$ determinant
\begin{equation*}
Q_{N}^{[m]}(z)=\gamma_{N} \left| \begin{array}{cccc}
         p_{N}^{[m]}(z) & p_{N+1}^{[m]}(z) & \cdots &  p_{2N-1}^{[m]}(z) \\
         p_{N-2}^{[m]}(z) & p_{N-1}^{[m]}(z) & \cdots &  p_{2N-3}^{[m]}(z) \\
        \vdots& \vdots & \vdots & \vdots \\
         p_{-N+2}^{[m]}(z) & p_{-N+3}^{[m]}(z) & \cdots &  p_{1}^{[m]}(z)
       \end{array}
 \right|,
\end{equation*}
where $p_{k}^{[m]}(z)= 0$ if $k<0$. When $m=1$, $Q_{N}^{[1]}(z)$ are the original Yablonskii-Vorob'ev polynomials $Q_{N}(z)$. When $m>1$, $Q_{N}^{[m]}(z)$ give higher members of this polynomial hierarchy. Root structures of the Yablonskii-Vorob'ev polynomial hierarchy have been studied in \cite{Clarkson2003-II}.

Through this Yablonskii-Vorob'ev polynomial hierarchy, the pattern of the $N$-th order NLS rogue wave (\ref{BilinearTrans2}) for large $a_{2m+1}$ and the other parameters $O(1)$ are asymptotically described as follows.

Far away from the origin, with $\sqrt{x^2+t^2}=O\left(|a_{2m+1}|^{1/(2m+1)}\right)$, there are $N_p$ fundamental (Peregrine) rogue waves, where \[ \label{Np}
N_p=\left[N(N+1)-N_{0}(N_{0}+1)\right]/2,
\]
and $N_{0}(N_{0}+1)/2$ is the multiplicity of the root zero in the polynomial $Q_{N}^{[m]}(z)$. This $N_0$ value is given explicitly by the equation
\begin{eqnarray}
&&N\equiv N_{0} \hspace{0.1cm} \mbox{mod} \hspace{0.1cm} (2m+1), \quad \text{or}   \label{N01}\\
&&N\equiv -N_{0}-1 \hspace{0.1cm} \mbox{mod} \hspace{0.1cm} (2m+1),   \label{N02}
\end{eqnarray}
under the restriction of $0\leq  N_{0} \leq m$. These Peregrine waves are $u_1(x-\hat{x}_{0}^{(k)}, t-\hat{t}_{0}^{\hspace{0.05cm}(k)})$, where
\begin{equation} \label{Pere}
u_1(x, t)=\left(1- \frac{4(1+2\textrm{i}t)}{1+4x^2+4t^2}\right)e^{\textrm{i}t},
\end{equation}
and their positions $(\hat{x}_{0}^{(k)}, \hat{t}_{0}^{\hspace{0.05cm}(k)})$ are given asymptotically by
\begin{eqnarray}
&&\hat{x}_{0}^{(k)}+\textrm{i}\hspace{0.05cm}\hat{t}_{0}^{\hspace{0.05cm}(k)}=\hat{z}_k, \label{x0t0}  \\
&&\hat{z}_{k}= z_{k}\left(-\frac{2m+1}{2^{2m}}a_{2m+1}\right)^{\frac{1}{2m+1}},   \label{zkhat}
\end{eqnarray}
with $z_k$ being any non-zero root of $Q_{N}^{[m]}(z)$. Thus, the rogue wave pattern formed by these Peregrine components has the same geometric shape as the root structure of the polynomial $Q_{N}^{[m]}(z)$, except for a dilation and rotation due to the multiplication factor on the right side of Eq. (\ref{zkhat}). In particular, the rotation-induced orientation of the rogue pattern is controlled by the phase of the complex parameter $a_{2m+1}$. 

In the neighborhood of the origin, where $\sqrt{x^2+t^2}=O(1)$, lies a $N_0$-th order rogue wave $u_{N_0}(x,t)$. This is a lower-order rogue wave given by Eq. (\ref{BilinearTrans2}), with its internal parameters $a_3, a_5, \cdots, a_{2N_0-1}$ the same as those in the original rogue wave. If $N_0=0$, then there will not be such a lower-order rogue wave at the origin.

Before proving these analytical results, let us compare these analytical predictions with true rogue wave patterns. For this purpose, we first show in Fig. 1 true rogue wave solutions (\ref{BilinearTrans2}) from the 3rd to 5th order, with large $a_3$, $a_5$, $a_7$ and $a_9$ in the first to fourth columns respectively. The specific large-parameter values in Fig. 1 are $a_3=(-60 \textrm{i}, -40 \textrm{i}, -30 \textrm{i})$ in the first column, $a_5=(-600 \textrm{i}, -400 \textrm{i}, -300 \textrm{i})$ in the second column, $a_7=(-1500 \textrm{i}, -2000 \textrm{i})$ in the third column, and $a_9=-3000 \textrm{i}$ in the last column, with the other parameters in each solution set as zero. It is seen that these rogue waves comprise a number of Peregrine waves forming triangular patterns for large $a_3$, pentagon patterns for large $a_5$, heptagon patterns for large $a_7$, and nonagon patterns for large $a_9$. In addition, some of these rogue waves contain a lower-order rogue wave at their centers. These rogue patterns resemble those plotted in \cite{KAAN2013} from Akhmediev breathers in the rogue-wave limit.

\small
\begin{figure}[htb]
\includegraphics[width=0.5\textwidth]{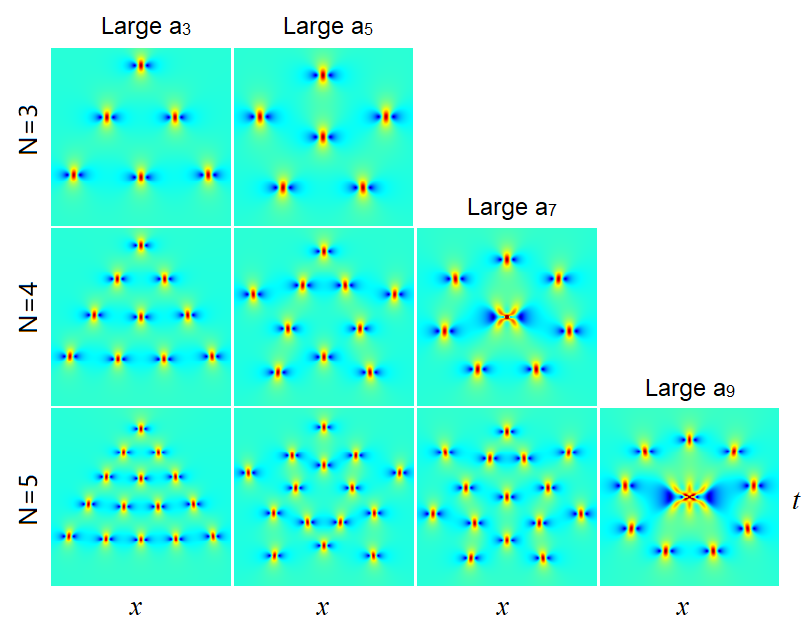}
\caption{NLS rogue wave patterns $|u_N(x,t)|$ of 3rd to 5th orders from true solutions (\ref{BilinearTrans2}) when one of the solution parameters is large and the other parameters set as zero.}
\end{figure}
\normalsize

Now, we compare these true rogue patterns with our analytical predictions. Our predicted solution $|u_{N}^{(p)}(x,t)|$ can be assembled into a simple formula,
\begin{equation} \label{upNLS}
\left|u_{N}^{(p)}(x,t)\right|=\left|u_{N_{0}}(x,t)\right| + \sum _{k=1}^{N_{p}}  \left(\left| u_1(x-\hat{x}_{0}^{(k)}, t-\hat{t}_{0}^{\hspace{0.05cm}(k)})\right| -1 \right),
\end{equation}
where $u_1(x,t)$ is the Peregrine wave given in (\ref{Pere}), and $u_{N_0}(x,t)$ is the lower-order rogue wave (\ref{BilinearTrans2}) with its internal parameters all zero, as inherited from the original rogue waves in Fig. 1. For the 5th order rogue waves, our predicted solutions for the same parameters of Fig. 1 are displayed in Fig. 2. These predicted patterns are strikingly similar to the true ones. We have also compared the predicted rogue waves of the 3rd and 4th orders to the true ones in Fig. 1, and the predictions are visually almost identical to the true solutions as well.

\small
\begin{figure}[htb]
\includegraphics[width=0.48\textwidth]{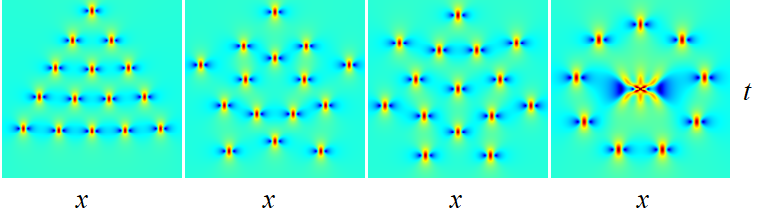}
\caption{Analytical predictions (\ref{upNLS}) for the 5-th order rogue waves in Fig. 1. The $x$ and $t$ intervals here are identical to those in Fig. 1, bottom row (for 5-th order). }
\end{figure}
\normalsize

It is illuminating to compare rogue patterns in Figs. 1 and 2 to root structures of the Yablonskii-Vorob'ev hierarchy $Q_{N}^{[m]}(z)$ reported in \cite{Clarkson2003-II}. Their geometric shapes are clearly the same, except for a rotation and dilation caused by the multiplication factor in our formula (\ref{zkhat}). This connection between rogue patterns and Yablonskii-Vorob'ev polynomials was never realized before to our best knowledge. 

Now, we sketch the proof of our analytical results on rogue patterns. Our proof is based on an asymptotic analysis of the rogue wave solution (\ref{BilinearTrans2}), or equivalently, the determinant $\sigma_n$ in Eq. (\ref{sigma_n}), in the large $a_{2m+1}$ limit, with the
other parameters being $O(1)$.

At large $(x,t)$ where $\sqrt{x^2+t^2}=O\left(|a_{2m+1}|^{1/(2m+1)}\right)$, the leading order term of $S_k(\textbf{\emph{x}}^{+}(n) +\nu \textbf{\emph{s}})$ is $S_k(\textbf{v})$, where $\textbf{v}=(x+\textrm{i}t, 0, \cdots, 0, a_{2m+1}, 0, \cdots)$. From the definition of Schur polynomials (\ref{Schurdef}), $S_k(\textbf{v})$ is given by
\begin{equation*}
\sum_{k=0}^{\infty}S_k(\textbf{v}) \epsilon^k
=\exp\left[(x+\textrm{i}t)\epsilon +a_{2m+1}\epsilon^{2m+1}\right].
\end{equation*}
Thus, it is related to the polynomial $p_{k}^{[m]}(z)$ in (\ref{pkmz}) as
\begin{equation*}
S_k(\textbf{v})=A^{k/(2m+1)}p_{k}^{[m]}(z),
\end{equation*}
where
\begin{equation} \label{Az}
A=-\frac{2m+1}{2^{2m}}a_{2m+1}, \hspace{0.3cm} z=A^{-1/(2m+1)}(x+\textrm{i}t).
\end{equation}
Using these formulae, we find that
\begin{equation*}
\det \left[S_{2i-j}(\textbf{\emph{x}}^{+}(n) +j \textbf{\emph{s}}) \right]\sim
\gamma_N^{-1}A^{\frac{N(N+1)}{2(2m+1)}}Q_{N}^{[m]}(z).
\end{equation*}
Similarly,
\begin{equation*}
\det \left[S_{2i-j}(\textbf{\emph{x}}^{-}(n) +j \textbf{\emph{s}}) \right]\sim
\gamma_N^{-1}\left(A^*\right)^{\frac{N(N+1)}{2(2m+1)}}Q_{N}^{[m]}(z^*).
\end{equation*}
Here, $S_k= 0$ when $k<0$.

To proceed further, we use determinant identities and the Laplace expansion to rewrite $\sigma_n$ in Eq. (\ref{sigma_n}) as \cite{OhtaJY2012}
\begin{eqnarray*}
&& \hspace{-0.8cm} \sigma_{n}=\sum
\det_{1 \leq i, j\leq N} \left[\frac{1}{2^{\nu_j}} S_{2i-1-\nu_j}(\textbf{\emph{x}}^{+}(n) +\nu_j \textbf{\emph{s}}) \right]  \nonumber \\
&&\hspace{0.33cm} \times \det_{1 \leq i, j\leq N}\left[\frac{1}{2^{\nu_j}}S_{2i-1-\nu_j}(\textbf{\emph{x}}^{-}(n) + \nu_j \textbf{\emph{s}})\right],
\end{eqnarray*}
where the summation is over all possible integers of $0\leq\nu_{1} < \nu_{2} < \cdots < \nu_{N}\leq 2N-1$. Since the highest order term of $a_{2m+1}$ in this $\sigma_n$ comes from the index choices of $\nu_{j}=j-1$, then
\begin{equation*}
\sigma_n \sim \alpha |a_{2m+1}|^{\frac{N(N+1)}{2m+1}} \left|Q_{N}^{[m]}(z)\right|^2,
\end{equation*}
where $\alpha$ is a $(m, N)$-related non-zero constant. This equation shows that $\sigma_1/\sigma_0\sim 1$, i.e., the solution $u(x,t)$ is on the unit background, except at or near $(x, t)$ locations $\left(\hat{x}_{0}, \hat{t}_{0}\right)$ where $z$ is a root of the polynomial $Q_{N}^{[m]}(z)$, and such $\left(\hat{x}_{0}, \hat{t}_{0}\right)$ locations are given by Eq. (\ref{x0t0}) in view of Eq. (\ref{Az}). Performing further asymptotic analysis, we can show that in the neighborhood of each of these $\left(\hat{x}_{0}, \hat{t}_{0}\right)$ locations, the solution $u(x,t)$ is asymptotically a Peregrine soliton (\ref{Pere}) centered at $\left(\hat{x}_{0}, \hat{t}_{0}\right)$.

Next, we analyze the solution $u(x,t)$ in the neighborhood of the origin, where $\sqrt{x^2+t^2}=O(1)$, when $a_{2m+1}$ is large and the other parameters being $O(1)$. In this case, we first rewrite the $\sigma_n$ determinant (\ref{sigma_n}) into a $3N\times 3N$ determinant \cite{OhtaJY2012}
\[ \label{3Nby3Ndet2}
\sigma_{n}=\left|\begin{array}{cc}
\textbf{O}_{N\times N} & \Phi_{N\times 2N} \\
-\Psi_{2N\times N} & \textbf{I}_{2N\times 2N} \end{array}\right|,
\]
where $\Phi_{i,j}=2^{-(j-1)} S_{2i-j}\left[\textbf{\emph{x}}^{+}(n) + (j-1) \textbf{\emph{s}}\right]$, and
$\Psi_{i,j}=2^{-(i-1)} S_{2j-i}\left[\textbf{\emph{x}}^{-}(n) + (i-1) \textbf{\emph{s}}\right]$.
Defining $\textbf{\emph{y}}^{\pm}$ to be the vector $\textbf{\emph{x}}^{\pm}$ without the $a_{2m+1}$ term, i.e., let
$\textbf{\emph{x}}^{\pm}=\textbf{\emph{y}}^{\pm}+(0, \cdots, 0, a_{2m+1}, 0, \cdots)$, we find that the Schur polynomials of $\textbf{\emph{x}}^{\pm}$ are related to those of $\textbf{\emph{y}}^{\pm}$ as
\begin{equation*}
S_{j}(\textbf{\emph{x}}^{\pm}+\nu\textbf{\emph{s}}) = \sum_{i=0}^{\left[\frac{j}{2m+1}\right]} \frac{a_{2m+1}^i}{i!} S_{j-(2m+1)i}(\textbf{\emph{y}}^{\pm}+\nu\textbf{\emph{s}}),
\end{equation*}
where $[a]$ represents the largest integer less than or equal to $a$. Using this relation, we express matrix elements of $\Phi$ and $\Psi$ in Eq. (\ref{3Nby3Ndet2}) through Schur polynomials $S_{k}(\textbf{\emph{y}}^{\pm}+\nu\textbf{\emph{s}})$ and powers of $a_{2m+1}$. Then, we perform a series of row operations to the $\Phi$ matrix so that certain high-power terms of $a_{2m+1}$ are eliminated. Afterwards, we keep only the highest power terms of $a_{2m+1}$ in each row of the remaining matrix. Similar column operations are also performed on the matrix $\Psi$. With these manipulations, we find that $\sigma_n$ is asymptotically reduced to
\[ \label{3Nby3Ndet3}
\sigma_{n}\sim \beta \left|\begin{array}{cc}
\textbf{O}_{N_0\times N_0} & \widehat{\Phi}_{N_0\times 2N_0} \\
-\widehat{\Psi}_{2N_0\times N_0} & \textbf{I}_{2N_0\times 2N_0} \end{array}\right|,
\]
where $\beta$ is a $(m, N)$-dependent constant,
\begin{eqnarray*}
&& \widehat{\Phi}_{i,j}=2^{-(j-1)} S_{2i-j}\left[\textbf{\emph{y}}^{+}(n) + (j-1+\nu_0) \textbf{\emph{s}}\right], \\
&& \widehat{\Psi}_{i,j}=2^{-(i-1)} S_{2j-i}\left[\textbf{\emph{y}}^{-}(n) + (i-1+\nu_0) \textbf{\emph{s}}\right],
\end{eqnarray*}
and $\nu_0$ is a certain integer. Finally, we notice that $S_j\left[\textbf{\emph{y}}^{\pm} + (\nu+\nu_0) \textbf{\emph{s}}\right]$ is related to $S_j\left(\textbf{\emph{y}}^{\pm} + \nu \textbf{\emph{s}}\right)$ through
\begin{equation*}
S_{j}\left[\textbf{\emph{y}}^{\pm} + (\nu+\nu_{0}) \textbf{\emph{s}}\right]=\sum_{i=0}^{\left[j/2\right]}S_{2i}(\nu_{0}\textbf{\emph{s}}) S_{j-2i}(\textbf{\emph{y}}^{\pm} +\nu\textbf{\emph{s}}).
\end{equation*}
Using this relation, $\sigma_n$'s determinant (\ref{3Nby3Ndet3}) can be reduced to one with $\nu_0=0$ in the above $\widehat{\Phi}$ and $\widehat{\Psi}$ matrices. Such $\sigma_n$ then gives a $N_0$-th order rogue wave, whose internal parameters $a_j$ are identical to those in the original $N$-th order rogue wave in (\ref{BilinearTrans2}).

As a small application of our analytical results, we explain the numerical observations in \cite{KAAN2011}. Under our bilinear rogue solution (\ref{BilinearTrans2}), a $N$-th order rogue wave exhibits a ring structure when $a_{2N-1}$ is large (see Fig. 1). In this case, $m=N-1$, and $N_0=N-2$ according to the formula (\ref{N02}). Then, our theory predicts that the center of this $N$-th order rogue wave is a $(N-2)$-the order rogue wave, surrounded by $N_p=2N-1$ Peregrine rogue waves on the ring. This is precisely what was observed in \cite{KAAN2011}.

Remarkably, this connection of rogue patterns to the root structures of the Yablonskii-Vorob'ev hierarchy holds for other integrable equations as well. To demonstrate, we consider the derivative NLS (DNLS) equation
\[ \label{GDNLS}
\textrm{i}u_t+\frac{1}{2}u_{xx}+\textrm{i}(|u|^2u)_x=0,
\]
which arises in very different physical situations from the NLS equation \cite{Kaup_Newell,KN_Alfven1}. Bilinear forms of rogue waves in this equation have been presented in \cite{YangDNLS2019} under the boundary conditions of $u(x, t) \to e^{-{\rm i}(1+\alpha) x-{\rm i}(\alpha^2-1)t/2}$ as $x, t\to \infty$, where $\alpha$ is a background wavenumber parameter. This bilinear rogue wave of $N$-th order also has free complex parameters $a_1, a_3, \dots, a_{2N-1}$, and we will normalize $a_1=0$ through shifts of $x$ and $t$ as for the NLS equation. Then, when one of these complex parameters, $a_{2m+1}$, is large, we find that far away from the origin, the rogue pattern also comprises $N_p$ fundamental DNLS rogue waves located at $(x, t)=\left(\hat{x}_{0}^{(k)}, \hat{t}_{0}^{\hspace{0.05cm}(k)}\right)$, where
\begin{eqnarray*}
&& \hat{x}_{0}^{(k)}= \frac{1}{\sqrt{\alpha}} \Real \left(\hat{z}_k\right) -\frac{\alpha-1}{\alpha}\Imag \left(\hat{z}_{k}\right), \quad \hat{t}_{0}^{\hspace{0.05cm}(k)}=\frac{1}{\alpha} \Imag\left(\hat{z}_{k}\right),
\end{eqnarray*}
$\hat{z}_k$ is as given in Eq. (\ref{zkhat}), and ``$\Real$, $\Imag$" represent the real and imaginary parts of a complex number.
When $\alpha=1$, these locations are identical to those in (\ref{x0t0}) for the NLS equation, which means that the DNLS rogue pattern formed by these fundamental DNLS rogue waves would be identical to those shown in Fig. 1 for the NLS equation. When $\alpha\ne 1$, the above
$\left(\hat{x}_{0}^{(k)}, \hat{t}_{0}^{\hspace{0.05cm}(k)}\right)$ locations are related to the root structure of the polynomial $Q_{N}^{[m]}(z)$ through a stretching (shear), in addition to dilation and rotation. In the neighborhood of the origin, the $N$-th order DNLS rogue wave also reduces to a lower $N_0$-th order DNLS rogue wave, with $N_0$ as given in Eqs. (\ref{N01})-(\ref{N02}). To verify these predictions, we set $\alpha=1$ and plot in Fig. 3 true DNLS rogue waves from the 3rd to 5th order, with one of the parameters large and the others set as zero. It is seen that these rogue patterns are indeed identical to those in Fig. 1 for the NLS equation. The only difference is the local shapes of DNLS fundamental rogue waves away from the origin and local shapes of
lower-order DNLS rogue wave at the origin. We have also compared these true DNLS rogue patterns with our analytical predictions (shown in Fig. 4). It is seen that the predictions closely match true DNLS rogue patterns.

\small
\begin{figure}[h]
\includegraphics[width=0.5\textwidth]{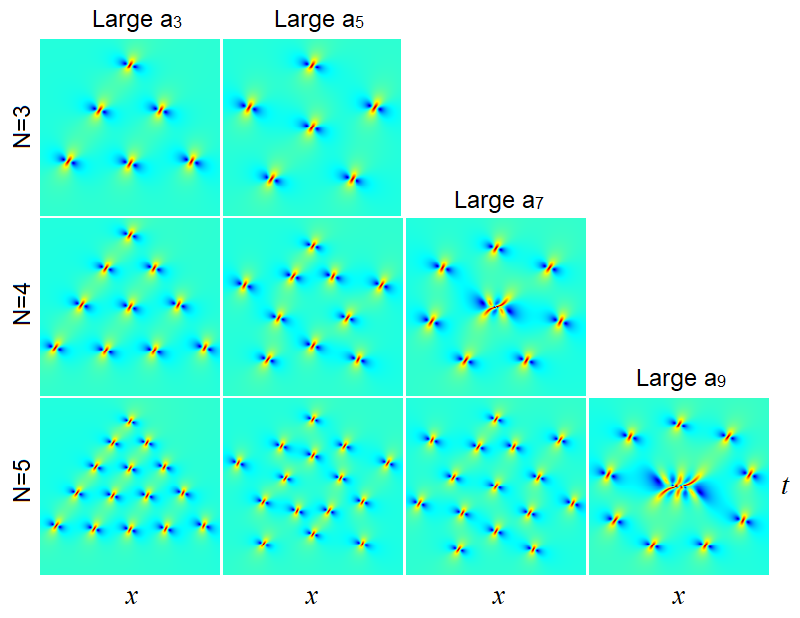}
\caption{True DNLS rogue patterns $|u_N(x,t)|$ of 3rd to 5th orders when one of the solution parameters is large and the other parameters set as zero. }
\end{figure}
\normalsize

We have also explored rogue patterns in other equations such as the Manakov equations
and the Boussinesq equation, and found that those patterns under large parameters are described
by the root structures of the Yablonskii-Vorob'ev polynomial hierarchy as well. Thus, rogue
waves in different equations develop universal patterns.

\small
\begin{figure}[h]
\includegraphics[width=0.48\textwidth]{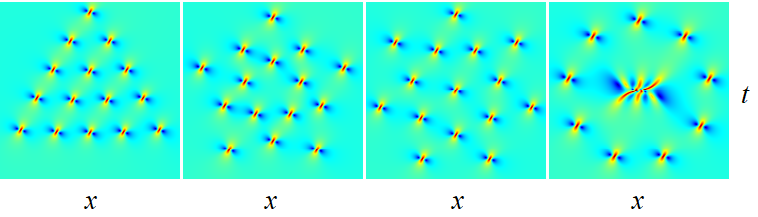}
\caption{Analytical predictions for 5-th order DNLS rogue waves of Fig. 3.}
\end{figure}
\normalsize

In summary, we have shown that universal rogue wave patterns appear in different physical systems, and these rogue patterns are analytically predicted by the root structures of the Yablonskii-Vorob'ev polynomial hierarchy. These results significantly improve our analytical understanding of rogue patterns, and they can be useful for the quantitative prediction of physical rogue events.

This work was supported in part by AFOSR (FA9550-18-1-0098) and NSF (DMS-1910282).

\end{document}